\documentclass[aps,prd,amsfonts,groupedaddress,showpacs,showkeys]{revtex4}

\usepackage{epsfig,tabularx,array,booktabs,calc,multirow,amsmath}

\newcommand{\beaa}{\begin{eqnarray*}} 
\newcommand{\enaa}{\end{eqnarray*}}
\newcommand{\bea}{\begin{eqnarray}}
\newcommand{\ena}{\end{eqnarray}}
\newcommand{\eq}{\begin{eqnarray}} 
\newcommand{\en}{\end{eqnarray}}
\newcommand{\ra}{\rangle}
\newcommand{\la}{\langle}

\begin{document}

\title{Strong and radiative decays of the scalars\\
$\mathbf{f_0(980)}$ and $\mathbf{a_0(980)}$ in a hadronic molecule approach.} 

\author{
Tanja Branz, 
Thomas Gutsche, 
Valery E. Lyubovitskij
\footnote{On leave of absence from the
Department of Physics, Tomsk State University,
634050 Tomsk, Russia} 
\vspace*{1.2\baselineskip}} 
\affiliation{Institut f\"ur Theoretische Physik,
Universit\"at T\"ubingen,
\\ Auf der Morgenstelle 14, D-72076 T\"ubingen, Germany
\vspace*{0.3\baselineskip}\\}

\date{\today}

\begin{abstract} 
We analyze the electromagnetic and strong decay properties of the light 
scalars $a_0(980)$ and $f_0(980)$ within a hadronic molecule interpretation. 
Both scalars are discussed within a covariant and gauge invariant model 
which also allows for finite size effects due to their spatially extended 
structure in the $K\bar K$-bound state picture. Allowing for $f_0-a_0$ mixing 
we also study its influence on the radiative decays 
$f_0/a_0\to \gamma\gamma$, $f_0/a_0\to \gamma\omega$, and 
$f_0/a_0\to \gamma\rho$ as well as the $\phi$ production of the $f_0$ and 
$a_0$. 
Furthermore, we apply our formalism to describe the strong $f_0\to\pi\pi$ 
and $a_0\to\pi\eta$ decay properties.

\end{abstract} 



\maketitle

\newpage
\section{Introduction}
Until now meson spectroscopy provides a valuable tool to explore the structure 
and properties of mesons and, extending the scope, to get further information 
on the confinement regime of strong interaction. During the last decade the 
meson mass spectrum showed a richer structure than might be expected from the 
constituent quark model, which decisively influenced our understanding of 
hadronic structure in the past. In particular, the structure issue of the 
lightest scalars has been under permanent discussion concerning mesonic 
structure beyond the quark-antiquark picture. There exist different approaches 
concerning the substructure of the $f_0(980)$ and its ``twin'', the 
$a_0(980)$, which range from 
$q\bar q$~\cite{Tornqvist:1995kr,vanBeveren:1998qe,%
Giacosa:2008xp} to tetraquark $q^2\bar q^2$ \cite{Jaffe:1976ig,%
Achasov:1981kh,Giacosa:2006rg} interpretations. In~\cite{Black:2007bp} 
the structure of the light scalar nonet including $f_0(980)$ was tested using 
radiative $\phi$ decays. The authors of Ref.~\cite{Black:2007bp} point out 
the difficulty to distinguish between the $q\bar q$ and the 
$q q \bar q \bar q$ picture for the light scalar mesons. A possible admixture 
between $\bar q q$ and $q q \bar q \bar q$ configurations for the low-lying 
scalar mesons has been considered in Ref.~\cite{Fariborz:2006ff} using the 
chiral approach. Both scalars are also discussed in a clustered version of 
the tetraquark configuration where the two quarks and antiquarks form a bound 
state of mesons - 
hadronic molecules~\cite{Weinstein:1982gc,Hanhart:2007wa,Kalashnikova:2005zz}. 
In addition, an isospin-violating mixture of the $f_0(980)$ and $a_0(980)$ 
mesons has been originally discussed in~\cite{Achasov:1979xc} and taken into 
consideration in~\cite{Close:2000ah,Achasov:2002hg,Hanhart:2007bd} which 
provides an interesting possibility to study its substructure. $f_0-a_0$ 
mixing is on the one hand motivated by their near degenerate masses, on the 
other hand by the mass gap between the nearby charged and neutral $K\bar K$ 
thresholds. A crucial check for theoretical considerations will be future 
experiments planning to investigate $f_0-a_0$ mixing (see e.g. 
Ref.~\cite{Wu:2007jh}). 

In the present paper we study the electromagnetic and strong decay properties 
of the $f_0(980)$ and $a_0(980)$ mesons which are assumed to be of a pure 
molecular meson structure, that is bound states of two kaons. We discuss the 
electromagnetic decays with the final states occupied by photons and massive 
vector mesons $S\to V\gamma,\,S\to \gamma\gamma$ and $\phi\to S\gamma$, where 
$S=f_0,a_0$ and $V=\rho,\omega$, as well as the strong 
$a_0/f_0\to \pi\pi/\pi\eta$ decay properties. 

For the description of the $f_0(980)$ and $a_0(980)$ as hadronic molecules 
we apply the theoretical framework developed in~\cite{Branz:2007xp} 
based on the use of the compositeness 
condition $Z=0$~\cite{Weinberg:1962hj,Efimov:1993ei} which implies that 
the renormalization constant of the hadron wave function is set equal to zero.
Note, that this condition was originally applied to the study of
the deuteron as a bound state of proton and neutron~\cite{Weinberg:1962hj}.
Then it was extensively used in the low-energy hadron
phenomenology as the master equation for the treatment of
mesons and baryons as bound states of light and heavy
constituent quarks (see Refs.~\cite{Efimov:1993ei,Faessler:2003yf}). 
In Refs.~\cite{Faessler:2007gv} the compositeness condition has 
been successfully used in the description of the recently discovered 
heavy mesons as hadronic molecules. In particular, within the mesonic 
bound state interpretation, the compositeness condition allows for 
a self-consistent determination of the coupling of the scalar mesons to 
their constituents. The advantage of our approach is that it has a clear 
and consistent mathematical structure with a minimal amount of free 
parameters. It also fulfills essential conditions such as covariance and 
gauge invariance, while allowing to include the spatially extended structure 
of the meson molecules and isospin-violating mixing effects. Here we generate 
the $f_0-a_0$ mixing due to the mass difference of intermediate charged and 
neutral kaon loops; this mechanism was proposed in~\cite{Achasov:1979xc} as 
the leading contribution to the $f_0-a_0$ mixing. Note that in our approach 
this mixing mechanism is naturally generated due to the coupling of $a_0$ 
and $f_0$ to its constituents - the kaons. 

The paper is organized as follows. Our framework is discussed in 
Sec.~\ref{sec:framework}. We derive the effective mesonic Lagrangian 
for the treatment of $f_0$ and $a_0$ as $K \bar K$ bound states 
(molecules) in Sec.~\ref{sec:molecule}. In Sec.~\ref{sec:mixing}
we discuss the modification of $f_0 K \bar K$ and $a_0 K \bar K$ couplings 
due to the $f_0-a_0$ mixing. In Sec.~\ref{sec:em} we include the 
electromagnetic interactions and discuss the diagrams contributing 
to the radiative decays of $f_0$ and $a_0$. Our results are presented in 
Sec.~\ref{sec:res}, which we also compare with other approaches and 
with experimental data. In Sec.~\ref{sec:summary} we present a short 
summary of our results.
\def\arraystretch{1.5} 
\section{Theoretical framework}\label{sec:framework}

\subsection{Molecular structure of the $f_0(980)$ and $a_0(980)$ mesons} 
\label{sec:molecule}

The theoretical framework we use for our analysis is based on the 
nonlocal strong Lagrangians~\cite{Branz:2007xp,Anikin:1995cf,Faessler:2003yf,%
Faessler:2001mr,Faessler:2007gv}
\eq
\begin{aligned}
{\cal L}_{f_0K\bar K}=&\frac{g_{f_0K\bar K}}{\sqrt2}f_0(x)\int dy\,
\Phi(y^2)\bar K\Big(x-\frac y2\Big)K\Big(x+\frac y2\Big)\,,\\
{\cal L}_{a_0K\bar K}=&\frac{g_{a_0K\bar K}}{\sqrt2}\vec a_0(x)\int dy\,
\Phi(y^2)\bar K\Big(x-\frac y2\Big)\vec\tau K\Big(x+\frac y2\Big)\,,
\end{aligned}\label{eq:L}
\en
describing the interaction between the kaon-antikaon bound state and 
its constituents. The kaon and scalar fields are collected in the kaon 
isospin doublets and the scalar meson triplet
\eq 
K=\left(\begin{array}{c}K^+\\K^0\end{array}\right)\,,
\quad\bar K=\left(\begin{array}{c}K^-\\\bar K^0\end{array}\right)
\quad\text{and}\quad\vec a_0=(a_0^+,\,a_0^0,\,a_0^-)\,.
\en 
The vector $\vec\tau=(\tau^+,\tau^0,\tau^-)$ is characterized by the 
Pauli matrices $\tau_{i=1,2,3}$, where 
$\tau^\pm=\frac{1}{\sqrt2}(\tau_1\pm i\tau_2)$ and $\tau^0=\tau_3$. 

Finite size effects are incorporated in our model by the correlation 
function $\Phi(y^2)$. Its Fourier transform $\widetilde\Phi(k_E^2)$ 
is directly related to the shape and size of the hadronic molecule and 
shows up as the form factor in the Feynman diagrams. 
Here, we employ a Gaussian form
\eq 
\Phi(y^2)=\int\frac{d^4k}{(2\pi)^4}\widetilde \Phi(-k^2)e^{-iky}\quad 
\text{with}\quad\widetilde\Phi(k_E^2)=\exp(-k_E^2/\Lambda^2)\,,
\en 
where the index $E$ refers to the Euclidean momentum space. The size 
parameter $\Lambda$ controls the spatial extension of the hadronic molecule 
and is varied around 1 GeV. In the special case of pointlike interaction, 
which we refer to as the local case, the correlation function $\Phi(y^2)$ 
is replaced by the delta function 
$\lim\limits_{\Lambda\to\infty}\Phi(y^2)=\delta^{(4)}(y)$. 

\begin{figure}[htbp]
 \includegraphics[trim= 0.5cm 22.5cm 0.0cm 1.50cm, clip,scale=0.6]{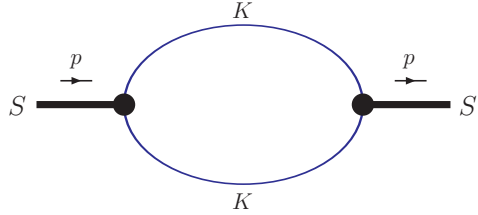}
 \caption{Mass operator of $S=f_0,a_0$.}
\label{fig1}
\end{figure}

The couplings to the constituent kaons, $g_{SK\bar K}$ with $S=f_0,a_0$, 
are determined self-consistently within our model by using the 
compositeness condition. It provides a method to fix the coupling strength 
between a bound state and its 
constituents~\cite{Weinberg:1962hj,Efimov:1993ei}; it therefore reduces the 
amount of free input parameters and also allows for a clear and 
straightforward determination of the decay properties. Note that this 
condition has also been used in the $K\bar K$ molecule approach 
in~\cite{Baru:2003qq,Hanhart:2007wa}. The coupling constant can be easily 
extracted from the definition of the field renormalization constant 
$Z_{f_0}$ which is set to zero
\eq
Z_{S}=1-\widetilde\Pi^\prime (M_{f_0}^2)=0\,.\label{sigma}
\en 
Here, $\widetilde\Pi^\prime(M_{f_0}^2)=\frac{g_{f_0K\bar K}^2}{(4\pi)^2}
\widetilde\Sigma^\prime(M_{f_0}^2)$ is the derivative of the mass operator 
\eq
\Sigma(p^2)=\int\frac{d^4k}{\pi^2i}\,\widetilde\Phi^2(-k^2)
\,S\Big(k+\frac p2\Big)S\Big(k-\frac p2\Big)\, 
\en 
shown in Fig.~\ref{fig1}. We stress that the Weinberg condition applies 
only to the bound states. 
In general, meson-loop diagrams are evaluated by using the free meson 
propagators given by 
\eq
i \, S_K(x-y) = \langle 0 | T \, K(x) \, K^\dagger(y)  | 0 \rangle
\ = \
\int\frac{d^4k}{(2\pi)^4i} \, e^{-ik(x-y)} \ S_K(k) \,,
\en
in case of scalar and pseudoscalar mesons, where 
\eq
S_K(k) = \frac{1}{M_K^2 - k^2 - i\epsilon}\,.
\en 
For vector and axialvector fields ($H^\ast = V, A$) we use
\eq 
iS_{H^\ast}^{\mu\nu}(x-y)=\left<0|TH^{\ast\,\mu}(x)
H^{\ast\,\nu\,\dagger} (y)|0\right> 
= \int\frac{d^4k}{(2\pi)^4i}\, e^{-ik(x-y)} S^{\mu\nu}_{H^\ast}(k)
\en
with
\eq 
S^{\mu\nu}_{H^\ast}(k)=
\frac{-g^{\mu\nu}+k^\mu k^\nu/M_{H^\ast}^2}{M_{H^\ast}^2-k^2-i\epsilon}\,.
\en 

\subsection{Inclusion of $f_0-a_0$ mixing}\label{sec:mixing}

\begin{figure}[htbp]
 \includegraphics[trim= 0.5cm 22.5cm 0.0cm 1.50cm, clip,scale=0.6]{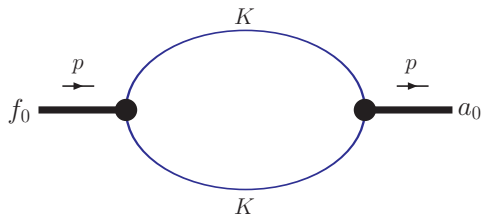}
 \caption{Leading contribution to $f_0-a_0$ mixing.}
\label{fig2}
\end{figure}

\begin{figure}[htbp]
 \includegraphics[trim= 0.7cm 12cm -1.2cm 6.0cm, clip,scale=0.7]{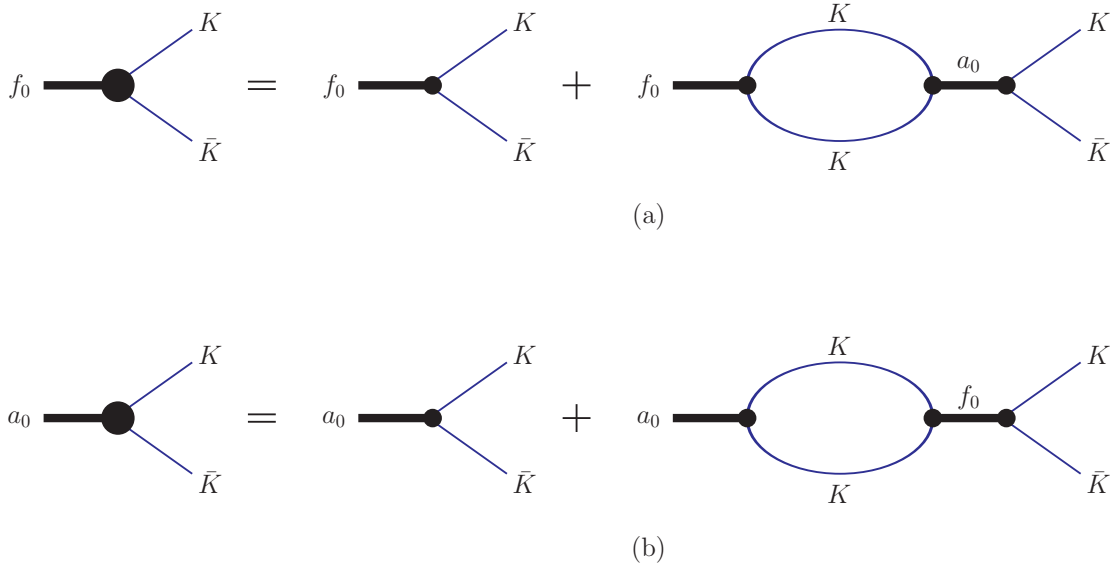}
 \caption{Renormalization of $f_0K\bar K$ and $a_0K\bar K$ couplings due 
 to $f_0-a_0$ mixing.}
\label{fig3}
\end{figure}

The isospin-violating mixture of the $f_0(980)$ and $a_0(980)$ mesons 
was originally discussed in~\cite{Achasov:1979xc} and also pursued later 
e.g. in Refs.~\cite{Close:2000ah,Achasov:2002hg,Hanhart:2007bd}. 
In particular, in Ref.~\cite{Achasov:1979xc} a model-independent result for 
the $f_0-a_0$ mixing amplitude was derived due to the subtraction of charged 
and neutral kaon-loop diagrams, which is valid for any value of 
external momenta. In our approach this mixing amplitude (see Fig.~\ref{fig2}) 
is naturally generated due to the coupling of $f_0$ and $a_0$ to their 
constituent kaons. In the following we restrict the calculation to this 
leading contribution of the $f_0-a_0$ mixing mechanism. The mixing effect 
leads to a renormalization of the $f_0/a_0$ couplings to the constituents. 
The modified $f_0 K \bar K$ and $a_0 K \bar K$ couplings 
are shown in Fig.~\ref{fig3}(a) and \ref{fig3}(b). For the $f_0$ and $a_0$ 
propagators we use the ones in the Breit-Wigner form: 
\eq 
D_S(p^2) = \frac{1}{M_S^2 - p^2 + i M_S \Gamma_S}  
\en 
where $\Gamma_S=\Gamma(M_S^2)$ is the total width of the
$S=f_0(a_0)$ meson. 

Note that the focus of the present considerations lie on the electromagnetic 
and strong scalar decay properties, where mixing modifies the coupling between 
the meson molecule and the $K\bar K$ constituents in the loop. In the 
following we show that this mixing effect is not so dramatic for isospin 
allowed transitions. However, for consistency 
we include such effects since the corresponding $f_0-a_0$ 
mixing insertions are naturally generated by our effective Lagrangian. 
We just stress that 
a more detailed theoretical analysis of the $f_0-a_0$ mixing effects was done 
in Refs.~\cite{Achasov:1979xc,Close:2000ah,Achasov:2002hg,Hanhart:2007bd}. 
A direct access to the mixing strength can be obtained from isospin-violating 
processes, such as the $J/\psi\to\phi f_0\to \phi a_0$ reaction, which is 
discussed in~\cite{Wu:2007jh}. 

\subsection{Inclusion of the electromagnetic interaction}\label{sec:em} 

\begin{figure}[htbp]
 \includegraphics[trim= 0.5cm 3cm 0.0cm 4.0cm, clip,scale=0.65]{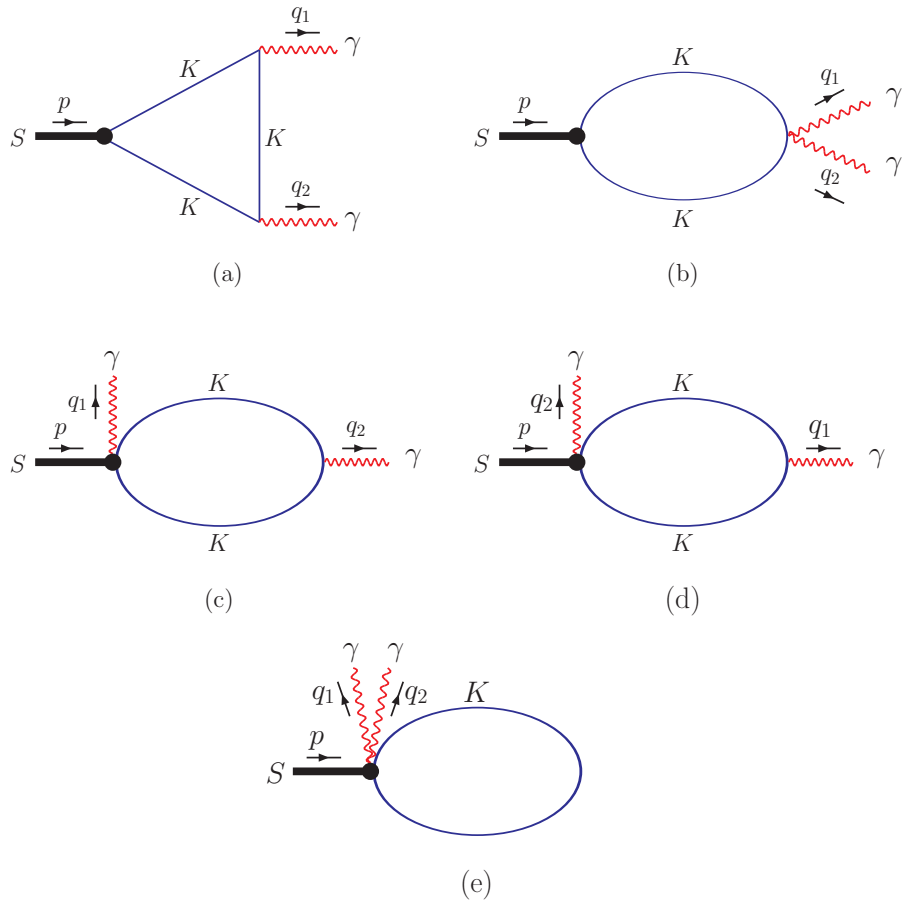}
 \caption{Diagrams contributing to the electromagnetic 
$f_0\rightarrow\gamma\gamma$ and $a_0\rightarrow\gamma\gamma$ decays.}
\label{fig4}
\end{figure}

\begin{figure}[htbp]
 \includegraphics[trim= 0cm 0cm 0cm 14.0cm, clip,scale=0.6]{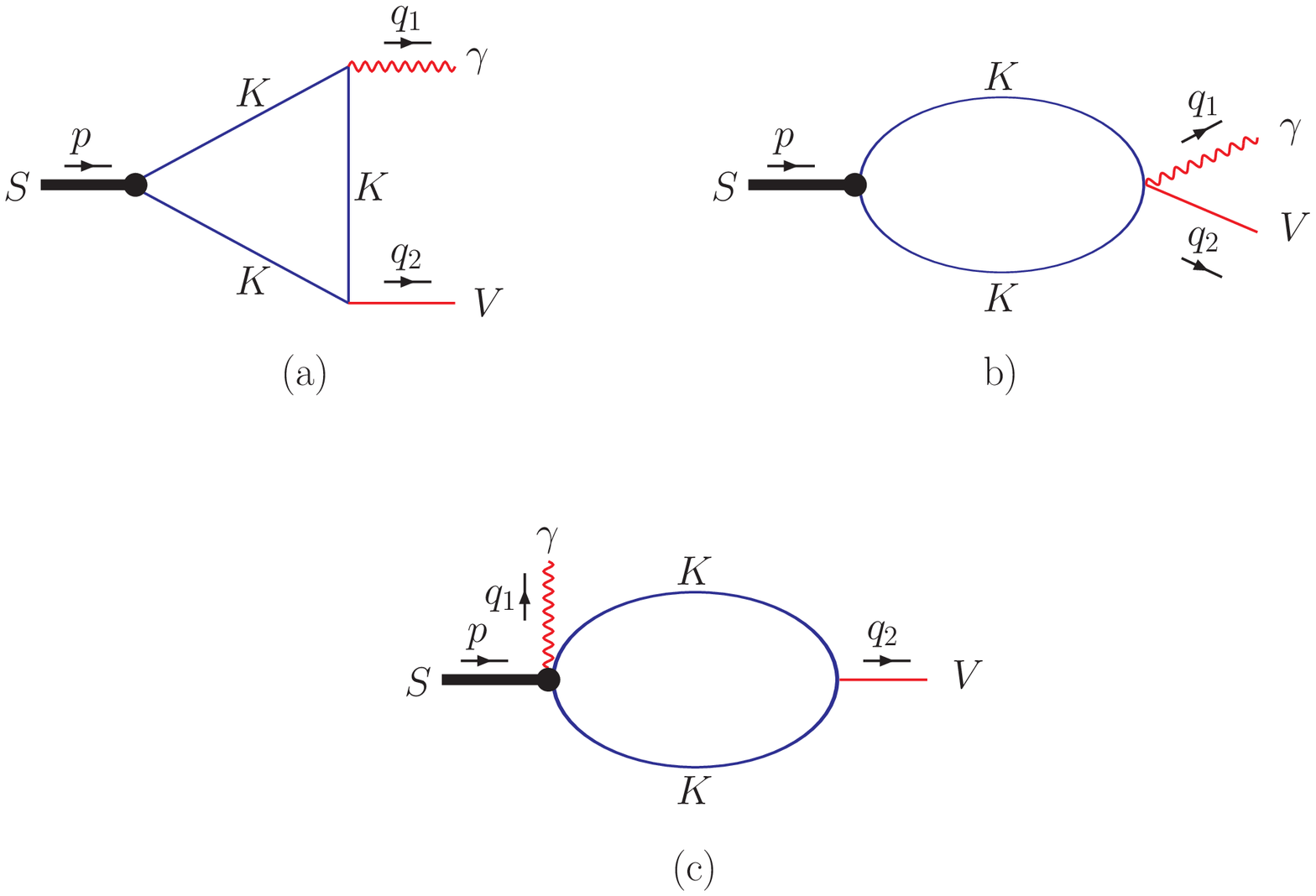}
 \caption{Diagrams describing the $S \to \gamma V$ decays.} 
\label{fig5}
\end{figure}

The electromagnetic interaction terms are obtained by minimal substitution 
$\partial^\mu K^\pm\to(\partial^\mu\mp ieA^\mu)K^\pm$ in the free 
Lagrangian ${\cal L}_K$ of charged kaons 
\eq 
{\cal L}_{K}=\partial_\mu K^+\partial^\mu K^--M_K^2K^+K^-
\en
and the Lagrangians which couple vector mesons and kaons
\eq
{\cal L}_{VK\bar K}=g_{\rho K\bar K}\vec \rho^{\,\mu} 
(\bar K \vec\tau \,i\partial_\mu K-K\vec\tau\, i\partial_\mu \bar K)+
(g_{\omega K\bar K}\omega^\mu+g_{\phi K\bar K}\phi^\mu)
(\bar K  \,i\partial_\mu K-K\, i\partial_\mu \bar K)\,.
\en 
The resulting electromagnetic interaction vertices are 
contained in the decay diagrams (a) and (b) of Figs. \ref{fig4} and 
\ref{fig5}. In the local limit, 
the decay amplitude would be completely described by these Feynman diagrams. 
In contrast, the nonlocal strong interaction Lagrangians  require special 
care in establishing gauge invariance. In doing so the charged fields are 
multiplied by exponentials \cite{Terning:1991yt} containing the 
electromagnetic field 
\eq 
K^\pm(y)\to e^{\mp ieI(y,x,P)}K^\pm(y)
\en 
with $I(y,x,P)=\int\limits_x^ydz_\mu A^\mu(z)$, which gives rise to 
the electromagnetic gauge invariant Lagrangian 
\eq
{\cal L}_{f_0K\bar K}^{GI} =\frac{g_{f_0K\bar K}}{\sqrt2}
f_0(x)\!\!\int\!dy\Phi(y^2)\big[e^{-ieI(x+\frac y2,x-\frac y2,P)}
K^+\big(\textstyle{x+\frac y2}\big)K^-\big(\textstyle{x-\frac y2}\big)
+K^0\big(\textstyle{x+\frac y2}\big)
\bar K^0\big(\textstyle{x-\frac y2}\big)\big]\,,
\en 
with a corresponding expression for the $a_0$ meson.
The interaction terms up to second order in $A^\mu$ are obtained 
by expanding ${\cal L}_{S K\bar K}^{GI}$ in terms of $I(y,x,P)$. 
Diagrammatically, the higher order terms give rise to nonlocal 
vertices with additional photon lines attached. The Feynman rules 
for these vertices have been already derived in~\cite{Faessler:2003yf}. 
Altogether, we obtain further graphs (Fig. \ref{fig4} (c), (d) and (e)) 
governing the two-photon decay and the diagram of Fig. \ref{fig5} (c) 
when massive vector mesons are involved. 
In a slightly modified form the diagrams of Fig. \ref{fig5} are also 
used to calculate the $\phi\to S\gamma$ 
decay~\cite{Close:1992ay,Kalashnikova:2004ta,Oller:2002na}. 
Quantitatively, the decay amplitude is dominantly characterized by the  
triangle diagram. The Feynman graphs containing contact vertices arising  
due to the nonlocality only give a minor contribution to the transition  
amplitude but are required in order to fully restore gauge invariance.  

The diagrams are evaluated by applying the technique developed 
in~\cite{Faessler:2003yf,Faessler:2001mr,Faessler:2007gv}, where each 
Feynman integral is separated into a part obeying gauge invariance and 
a remainder term. The remainder terms of each graph cancel each other 
in total and only the gauge invariant structure of the decay matrix 
element is left. The matrix element can therefore be written by a linear 
combination of the form factors $F(p^2,q_1^2,q_2^2)$ and 
$G(p^2,q_1^2,q_2^2)$ of the respective decay
\eq 
{\cal M}^{\mu\nu} = e^2 \Big( F(p^2,q_1^2,q_2^2) b^{\mu\nu}
                            + G(p^2,q_1^2,q_2^2) c^{\mu\nu} \Big)\,,
\en 
where the tensor structures are given by 
\bea
b^{\mu\nu}&=&g^{\mu\nu}(q_1q_2)-q_1^\mu q_2^\nu\label{eq:b}\\
c^{\mu\nu}&=&g^{\mu\nu}q_1^2q_2^2+q_1^\mu q_2^\nu(q_1 q_2)
-q_1^\mu q_1^\nu q_2^2-q_2^\mu q_2^\nu q_1^2\,.\nonumber
\ena
Here, $p$ and $q_1$ are the four-momenta of the scalar meson and photon; 
$q_2$ is the momentum of the vector meson or second photon depending on 
the respective decay. 

Since in the transition processes we deal with at least one real photon, 
the second part of ${\cal M}^{\mu\nu}$ proportional to $c^{\mu\nu}$ 
vanishes. The decay constant is therefore characterized by the form factor 
$F$ which is obtained by evaluating the Feynman integrals for on-shell 
initial and final states, where $V=\rho,\,\omega,\,\phi,\,\gamma$ 
represents the vector particle appropriate for the respective decay. 
In order to allow for $f_0-a_0$-mixing, we use $g_{f_0K^+K^-}$ and 
$g_{a_0K^+K^-}$ to compute the couplings characterizing the 
electromagnetic decays
\bea
g_{S\gamma \gamma}&\equiv& F_{S\gamma \gamma}(M_{S}^2,0,0)=
\frac{2}{(4\pi)^2} \frac{G_{SK\bar K}}{\sqrt2} 
I_{S\gamma \gamma}(M_{S}^2,0,0)\nonumber\\
g_{S\gamma V}&\equiv& F_{S\gamma V}(M_{S}^2,0,M_V^2)=
\frac{2}{(4\pi)^2}g_{VK\bar K}\frac{G_{SK\bar K}}{\sqrt2} 
I_{S\gamma V}(M_{S}^2,0,M_V^2)
\label{eq:F}\\
g_{\phi S\gamma}&\equiv& F_{\phi S\gamma}(M_\phi^2,M_{S}^2,0)=
\frac{2}{(4\pi)^2}g_{\phi K\bar K}\frac{G_{SK\bar K}}{\sqrt2} 
I_{\phi S\gamma}(M_\phi^2,M_{S}^2,0)\,,\nonumber
\ena
where $I$ denotes the loop integrals and $G_{f_0K\bar K}$ and $G_{a_0K\bar K}$ 
are the dressed couplings due to $f_0-a_0$-mixing. The explicit expressions 
for the loop integrals $I$ are given in Appendix A.
The issue of gauge invariance is considered in more detail in Appendix B 
and in the case of the two-photon decay 
in~\cite{Faessler:2003yf,Branz:2007xp}. 
In~\cite{Branz:2007xp} we also considered nontrivial $K\bar K \gamma$ 
interaction vertices, where these effects are absorbed in monopole form 
factors $F_{K\bar K\gamma}(Q^2)=\frac{1}{1+Q^2/\Lambda_{K\bar K\gamma}^2}$ 
depending on the photon momentum $Q^2$. However, this photon form factor 
does not influence the decay properties when dealing with real photons as 
in the present considerations.

\subsection{Strong decays}\label{sec:strong}

In order to calculate the strong decays of the $f_0$ and $a_0$ mesons 
we proceed in analogy with the computation of the $f_0\to \pi\pi$ decay 
in~\cite{Branz:2007xp}. In the present paper we extend the formalism by 
including the $a_0\to \pi\eta$ decay and, additionally, by considering mixing 
between both scalars. 

\begin{figure}[htbp]
\includegraphics[trim =0cm 16cm 0cm 7cm,clip,scale=0.7]{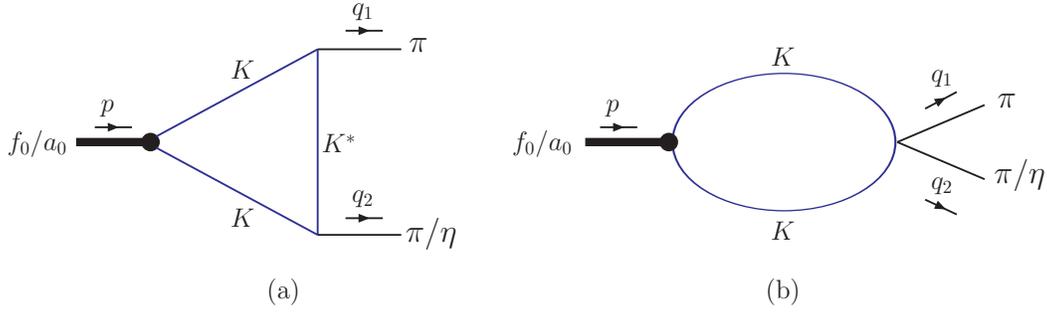}
\caption{Diagrams contributing to the strong decays}
\label{fig6}
\end{figure}

According to the interaction Lagrangians 
\eq
{\cal L}_{K^\ast K\pi}&=&
\frac{g_{K^\ast K\pi}}{\sqrt2}{K^\ast_\mu}^\dagger 
\vec \pi \vec \tau \,i{\partial^{^{^{\!\!\!\!\leftrightarrow}}}}^{\mu} K 
+ h.c\,,
\\
{\cal L}_{ K^\ast K\eta}&=&\frac{g_{K^\ast K\eta}}{\sqrt2}{K^\ast_\mu}^\dagger 
\eta  \,i{\partial^{^{^{\!\!\!\!\leftrightarrow}}}}^{\mu} K+h.c. 
\en
the final-state interaction effect in the $t$-channel proceeds via $K^\ast$ 
exchange (see Fig. \ref{fig6} (a)), where the massive vector meson is 
described by the antisymmetric tensor field $W_{\mu\nu} = - W_{\nu\mu}$. 
Therefore, the phenomenological Lagrangian which generates the contributing 
meson-loop diagrams is characterized by the Lagrangian
\eq 
{\cal L}_{W}(x) 
= - \frac{1}{2} \, \la \, \nabla^\sigma W_{\sigma\mu} 
\nabla_\nu W^{\nu\mu} + i G_V W_{\mu\nu} [u^\mu u^\nu] \, \ra \,,
\en 
which involves vector mesons in the tensorial 
representation~\cite{Gasser:1983yg,Ecker:1988te,Ecker:1989yg}. 
By using low-energy theorems $G_V$ can be expressed through the 
leptonic decay constant $G_V=F/\sqrt2$. The $K^\ast$ propagators 
in vector representation $S^V_{K^\ast;\mu\nu,\alpha\beta}(x)$ and 
tensorial description $S^W_{K^\ast;\mu\nu,\alpha\beta}(x)$ differ 
by a term which is reflected in a second diagram containing an 
explicit four meson vertex (see Fig. \ref{fig6} (b))
\eq 
S^W_{K^\ast;\mu\nu,\alpha\beta}(x)=
S^V_{K^\ast;\mu\nu,\alpha\beta}(x)+\frac{i}{M_{K^\ast}^2}
[g_{\mu\alpha}g_{\nu\beta}-g_{\mu\beta}g_{\nu\alpha}]\delta^4(x)\,.
\en 

Note that we include the interaction of four pseudoscalar mesons at leading 
$O(p^2)$ order in the chiral expansion given by chiral perturbation 
theory (ChPT)~\cite{Weinberg:1978kz,Gasser:1983yg}: 
\eq
{\cal L}_U(x) = \frac{F^2}{4} \la \, D_\mu U(x) D^\mu U^\dagger(x) + 
\chi U^\dagger(x) + \chi^\dagger U(x) \, \ra\,,
\en 
which leads to the four meson $\pi\pi K\bar K$ interaction vertex.
Inclusion of e.g. scalar resonances 
in the $s$-channel is of higher order, $O(p^4)$. In the $t$-channel we 
include the important vector meson exchange which also is of higher order, 
$O(p^4)$, but is important for the inclusion of final-state interactions. 
Here we use the standard notations of ChPT.   
The fields of pseudoscalar mesons are collected in the chiral matrix 
$U = u^2 = \exp(i\sum_i \phi_i \lambda_i/F)$ with $F = 92.4$ MeV 
being the leptonic decay constant and $D_\mu$ is the covariant derivative 
acting on the chiral field. Furthermore $\chi = 2 B {\cal M} + \cdots$, 
where $B$ is the quark vacuum condensate parameter 
$B = - \la 0|\bar u u|0 \ra/F^2 = - \la 0|\bar d d|0 \ra/F^2$ and 
${\cal M} = {\rm diag}\{\hat m, \hat m, M_S\}$ is the mass matrix of current 
quarks with $\hat m = (M_u + M_d)/2$. In the leading order of the chiral 
expansion the masses of pions and kaons are given by
$M_{\pi}^2=2 \hat m B, \hspace*{.2cm} M_{K}^2=(\hat m + M_S) B\,.$
In summary, second order ChPT gives rise to a second diagram being of the 
same structure as graph b) but opposite in sign. Therefore, the triangle 
diagram a) gives the dominant contribution to the decay amplitude.

The couplings for the strong decays are defined by
\eq 
g_{f_0\pi\pi}&=& g_{f_0\pi^+\pi^-}=
2g_{f_0\pi^0\pi^0}=G(M_{f_0}^2,M_\pi^2,M_{\pi}^2)\\
g_{a_0\pi\eta}&=&G(M_{a_0}^2,M_\pi^2,M_{\eta}^2)\,,
\en
where, in the case of the two-pion decay, we have to consider the ratio 
between the charged and neutral decay modes.
Here, $G(p^2,q_1^2,q_2^2)$ is the structure integral of the 
$f_0 \to \pi \pi$ and $a_0\to\pi\eta$ transitions, which are conventionally 
split into the two terms $G^{(a)}(p^2,\,q_1^2,\,q_2^2)$ and 
$G^{(b)}(p^2,\,q_1^2,\,q_2^2)$. They refer to the 
contributions of the diagrams of Figs.~\ref{fig6}~(a) and \ref{fig6}~(b), 
respectively, with 
\eq 
G(p^2,q_1^2,q_2^2)&=& G^{(a)}(p^2,q_1^2,q_2^2) + 
                      G^{(b)}(p^2,q_1^2,q_2^2)\,, 
\en 
where
\eq
G(p^2,q_1^2,q_2^2)=\dfrac{G_{SK\bar K}}{\sqrt2}\cdot 
\big(I(M_{K^\pm}^2,p^2,q_1^2,q_2^2)+I(M_{K^0}^2,p^2,q_1^2,q_2^2)\big)\,,
\en
and $I(M_{K}^2,p^2,q_1^2,q_2^2)$ denotes the contributions from the 
intermediate charged and neutral kaons.

The expressions for the decay widths are finally given by 
\eq
\Gamma(f_0\to\pi\pi)&=&\Gamma_{f_0\pi^+\pi^-}+\Gamma_{f_0\pi^0\pi^0} 
= \frac{3}{2}\Gamma_{f_0\pi^+\pi^-}=\frac{3}{32\pi}
\frac{g_{f_0\pi\pi}^2}{M_{f_0}}\sqrt{1-\frac{4M_{\pi}^2}{M_{f_0}^2}}\,,
\label{eq: widthstr}\\
\Gamma(a_0\to\pi\eta)&=&\frac{1}{16\pi}\frac{g_{a_0\pi\eta}^2}{M_{a_0}}
\frac{\lambda^{1/2}(M_{a_0}^2,M_{\pi}^2,M_\eta^2)}{M_{a_0}^2}\,,
\en 
with the K\"allen-function $\lambda(x,y,z)=x^2+y^2+z^2-2xy-2xz-2yz$.

\section{Results}\label{sec:res}
 
In this section we present our predictions for the electromagnetic and 
strong decay properties of the scalars $f_0$, $a_0$ and its sensitivity 
to finite size as well as mixing effects due to isospin-violation.

For all the numerical determinations we explicitly use the charged and 
neutral kaon masses $M_{K^\pm} = 493.677$ MeV and $M_{K^0} = 497.648$ MeV, 
since we consider isospin breaking effects. 

For the coupling constants between the hadronic molecules and the 
constituent kaons we obtain 
\eq
\dfrac{G_{f_0K\bar K;L}}{\sqrt2}&=&2.87\text{ GeV (local), 
\hspace*{0.5cm}}\dfrac{G_{f_0K\bar K}}{\sqrt2}=3.06\text{ GeV 
($\Lambda=1$ GeV)}\,,\nonumber\\
\dfrac{G_{a_0K\bar K;L}}{\sqrt2}&=&2.44\text{ GeV (local), 
\hspace*{0.5cm}}\dfrac{G_{a_0K\bar K}}{\sqrt2}=2.55 
\text{ GeV ($\Lambda=1$ GeV)}\,,
\en 
where the index $L$ refers to the local case.
For the computation of the radiative decay properties we use the vector 
meson masses quoted in \cite{Amsler:2008zz}
\eq
M_\rho&=&0.7755 \text{ GeV}\,,\nonumber\\
M_\omega&=&0.78265\text{ GeV}\,,\\
M_\phi&=&1.02\text{ GeV}\,. \nonumber
\en
The respective $g_{VK\bar K}$ and $g_{\omega K\bar K}$ couplings are fixed 
using the $SU(3)$ symmetry constraint:
\eq
g_{\rho K\bar K}=g_{\omega K\bar K}=\frac{g_{\phi K\bar K}}{\sqrt2}
=\frac{g_{\rho\pi\pi}}{2}=3
\en
with $g_{\rho K\bar K}=6$ extracted from data on the 
$\rho\to\pi^+\pi^-$ decay. Note, that the $SU(3)$ value for the 
$g_{\phi K\bar K}$ coupling (4.24) is close to the one predicted by 
data on the $\phi\to K^+K^-$ decay. In particular, using the formula 
for the $\phi\to K^+K^-$ decay width
\eq
\Gamma(\phi\to K^+K^-)=\frac{g_{\phi K\bar K}^2}{48\pi}
M_\phi\Big(1-\frac{4M_K^2}{M_\phi^2}\Big)^{3/2}
\en
and the central value for $\Gamma(\phi\to K^+K^-)$=2.10 MeV we deduce 
$g_{\phi K\bar K}$=4.48. The expressions for the electromagnetic decay 
widths are given by
\eq
\Gamma_{S\gamma\gamma}&=&\frac{\alpha^2\pi}{4}M_{S}^3g_{S\gamma\gamma}^2\,, 
\nonumber\\
\Gamma_{S\gamma\rho/\omega}&=&\frac{\alpha}{8}
\frac{(M_{S}^2-M_\rho^2)^3}{M_{S}^3}g_{S\gamma\rho/\omega}^2\,,\\
\Gamma_{\phi S\gamma}&=&\frac{\alpha}{24}
\frac{(M_{\phi}^2-M_S^2)^3}{M_{S}^3}g_{\phi S\gamma}^2\,, \nonumber
\en 
where the coupling constants describing the radiative decays are related 
to the form factor $F$ as described in (\ref{eq:F}). 
 
Within our hadronic molecule approach we obtain for the two-photon decay 
width of the $f_0(980)$ 
\eq 
\Gamma(f_0\to\gamma\gamma)&=&0.29\text{ (0.29) keV (local)}\,, 
\nonumber\\
\Gamma(f_0\to\gamma\gamma)&=&0.24\text{ (0.25) keV ($\Lambda=1$ GeV)}\,.
\en
The value in brackets refers to the corresponding value when neglecting 
$f_0-a_0$ mixing effects.
The sensitivity of the $f_0\to\gamma\gamma$ decay properties on finite 
size effects has been intensely studied in \cite{Branz:2007xp}, 
even in the case of virtual photons, and leads to a variation of 
$\Gamma(f_0\to\gamma\gamma)$ with the result
\eq 
\Gamma(f_0\to\gamma\gamma)=0.21\text{ keV ($\Lambda=$0.7 GeV)}-\;
0.26\text{ keV ($\Lambda=1.3$ GeV)} \,. 
\en 
In Tables~\ref{tab:em1} and~\ref{tab:em2} we draw the comparison with data 
and other approaches, respectively. The $f_0\to\gamma\gamma$ width predicted 
by our model matches the range of values currently deduced by the experiment. 

\begin{table}[htbp]
\caption{Electromagnetic decay width $f_0(980)\to\gamma\gamma$: 
experimental data.} 
\label{tab:em1}
\begin{tabularx}
{\linewidth}{lcccc}
\toprule\hline
Experiment &~\cite{Amsler:2008zz}&~\cite{Mori:2006jj}
           &~\cite{Marsiske:1990hx}&~\cite{Boyer:1990vu}\\ 
\hline
  $\Gamma(f_0\rightarrow\gamma\gamma)\quad$ [keV]
& $\quad 0.29^{+0.07}_{-0.09}\quad$
& $\quad 0.205^{+0.095\,+0.147}_{-0.083\,-0.117}\quad$ 
& $\quad 0.31\pm0.14\pm0.09\quad$&$\; 0.29\pm0.07\pm0.12$\\
\hline
\bottomrule
\end{tabularx}
\end{table}

\begin{table}[htbp]
\caption{Electromagnetic decay width $f_0(980)\to\gamma\gamma$: 
theoretical approaches.}
\label{tab:em2} 
\begin{tabularx}{\linewidth}{lXcXcXcXcXcXcXc}
\toprule\hline
Reference &&\cite{Efimov:1993ei}&&\cite{Anisovich:2001zp}
          &&\cite{Scadron:2003yg}&&\cite{Schumacher:2006cy}
          &&\cite{Achasov:1981kh}&&\cite{Oller:1997yg}
          &&\cite{Hanhart:2007wa}\\\hline
Meson structure&&$(q\bar q)$&&$(q\bar q)$&&$(q\bar q)$
          &&$(q\bar q)$&&$(q^2\bar q^2)$&&(hadronic)&&(hadronic)\\\hline
$\Gamma(f_0\rightarrow\gamma\gamma)\quad$ [keV] 
          &&$0.24$&&$0.28^{+0.09}_{-0.13}$&&$0.31$&&$0.33$
          &&$0.27$&&$0.20$&&$0.22 \pm 0.07$\\
\hline\bottomrule
\end{tabularx}
\end{table}

For the two-photon decay of the $a_0$ meson our results lie between 
 \eq 
\Gamma(a_0\to\gamma\gamma)&=&0.26\text{ (0.23) keV (local)}\,, 
\nonumber\\
\Gamma(a_0\to\gamma\gamma)&=&0.21\text{ (0.19) keV ($\Lambda=1$ GeV)}\,.
\en 
where again results without mixing are put in parentheses. By considering 
in addition the $f_0-a_0$ mixing contributions our estimates are in good 
agreement with the experimental result $0.3\pm0.1$ keV of 
Crystal Barrel~\cite{Amsler:1997up}. Finite size effects play a comparable 
role as $f_0-a_0$ mixing since the variation of $\Lambda$ from 0.7 GeV to 
1.3 GeV changes $\Gamma(a_0\to\gamma\gamma)$ by 
\eq 
\Gamma(a_0\to\gamma\gamma)=0.16\text{ keV ($\Lambda=0.7$ GeV) - 0.21 keV 
($\Lambda=1.3$ GeV)}\,.
\en 
The decay widths obtained in other approaches are combined in 
Table~\ref{tab:em3} and show a large discrepancy even for models with the 
same structure assumptions.
\begin{table}[htbp]
\caption{Electromagnetic decay width $a_0(980)\to\gamma\gamma$: 
theoretical approaches.}
\label{tab:em3}
\begin{tabularx}{\linewidth}{lXcXcXcXcXcXc}
\toprule\hline
Reference &&\cite{Anisovich:2001zp}&&\cite{Barnes:1985cy}
&&\cite{Achasov:1981kh}&&\cite{Oller:1997yg}\\\hline
Meson structure&&$(q\bar q)$&&$(q\bar q)$&&$(q^2\bar q^2)$
&&(hadronic)\\\hline
$\Gamma(a_0\rightarrow\gamma\gamma)$ [keV]&&$0.3^{+0.11}_{-0.10}$
&&1.5&&0.27&&0.78\\
\hline\bottomrule
\end{tabularx}
\end{table}

The radiative $\phi$ decay widths calculated in the local 
limit within the framework of our formalism are given by\eq
\Gamma(\phi\to f_0\gamma)=0.63 \text{ keV}\,,\nonumber\\
\Gamma(\phi\to a_0\gamma)=0.41\text{ keV}\,,
\en
where without mixing we obtain $\Gamma(\phi\to f_0\gamma)$=0.64 keV 
and $\Gamma(\phi\to a_0\gamma)$=0.37 keV. 

Our result for the $\phi\to f_0\gamma$ decay overestimates the value 
quoted by PDG (2007) \cite{Yao:2006px}, where the branching ratio 
$\Gamma(\phi\to f_0\gamma)/\Gamma_{total}=(1.11\pm0.07)\cdot10^{-4}$ 
yields $\Gamma(\phi\to f_0\gamma)$=0.44-0.51 keV. 
In the 2008 edition of PDG \cite{Amsler:2008zz}, the ratio is increased 
$\Gamma(\phi\to f_0\gamma)/\Gamma_{total}=(3.22\pm0.19)\cdot10^{-4}$ 
which gives 1.28-1.47 keV for the $\phi\to a_0\gamma$ decay width. 
However our results lie within the error bars of the CMD-2 
data \cite{Akhmetshin:1999di} $\Gamma(\phi\to f_0\gamma)=0.48-2.00$ keV. 

The decay width for the $\phi\to a_0\gamma$ decay slightly overestimates 
the PDG (2008) average value 0.3-0.35 keV 
($\Gamma(\phi\to a_0\gamma)/\Gamma_{total}=(0.76\pm0.06)\cdot10^{-4}$) 
but is in agreement with the experimental data of \cite{Achasov:2000ku} 
predicting 0.30-0.45 keV for $\Gamma(\phi\to a_0\gamma)$.

Because of the self-consistent determination of the $g_{a_0K\bar K}$ 
coupling constant our result in the case of the $a_0$ production is 
smaller than the width $\Gamma(\phi\to a_0\gamma)$ quoted 
in \cite{Close:1992ay,Kalashnikova:2005zz}, but we have quite good 
agreement with the predictions for the $\phi$-production of the $f_0$.

For the decays involving $\rho$ and $\omega$ mesons we predict:
\eq
\Gamma(f_0\to\rho\gamma)&=&\text{7.93 (8.09) keV (local) 
\hspace*{0.4cm}and\hspace*{0.4cm}  7.44 (7.58) keV ($\Lambda=1$ GeV)}\,,
\nonumber\\
\Gamma(f_0\to\omega\gamma)&=&\text{7.43 (7.57) keV (local) 
\hspace*{0.4cm}and\hspace*{0.4cm}  6.99 (7.12) keV ($\Lambda=1$ GeV)}\,,\\
\Gamma(a_0\to\rho\gamma)&=&\text{7.94 (7.18) keV (local) \hspace*{0.4cm}
and\hspace*{0.4cm} 7.29 (6.59) keV ($\Lambda=1$ GeV)}\,,\nonumber\\
\Gamma(a_0\to\omega\gamma)&=&\text{7.47 (6.76) keV (local) 
\hspace*{0.4cm}and\hspace*{0.4cm} 6.88 (6.22) keV ($\Lambda=1$ GeV)}\,.
\nonumber
\en
The deviations from the predicted widths of Ref.~\cite{Kalashnikova:2005zz} 
for the $a_0/f_0\to\gamma\rho/\omega$ decays arise because of different 
assumptions for the scalar masses and couplings. In~\cite{Nagahiro:2008mn} 
the decay width $a_0\to\gamma\rho/\omega$ calculated within the framework 
of a chiral unitarity approach is larger than our result because of the 
additional inclusion of vector mesons in the loop diagrams. 

In Appendix C our full results for the radiative decays of the neutral 
scalars $a_0$ and $f_0$ are collected in Table~\ref{tab1}. In the nonlocal 
case we have chosen $\Lambda$=1 GeV. For comparison we also indicate the 
decay properties when mixing effects are neglected. For simplicity the 
calculations for the $\phi$-decay are restricted to the local limit.

In summary, our results for the 
electromagnetic $f_0$ and $a_0$ decay properties are in quite good 
agreement with present experimental data. Therefore, the hadronic molecule 
approach is suitable to describe radiative $f_0$ and $a_0$ decays. However, 
other structure components besides the $K\bar K$ configuration can possibly 
be realized. Therefore, current data do not allow any definite and final 
conclusion concerning the substructure of the scalar mesons since 
calculations based on other approaches give similar results and even overlap 
with each other as demonstrated in Tables~\ref{tab:em2} and~\ref{tab:em3}. 

A further step forward would be a more precise experimental determination 
of the decay properties but also of the $f_0-a_0$ mixing strength to shed 
light on the isospin-violating mixing mechanisms. A possible access to mixing 
is given by the ratio between charged and neutral 
$a_0$ meson decays since the coupling to the charged $a_0^\pm$ mesons is 
not affected by mixing. 

In the numerical computations of the strong $f_0\to\pi\pi$ and 
$a_0\to\pi\eta$ decays we restrict to the charged pion mass 
($M_\pi\equiv M_{\pi^\pm}$139.57 MeV) but consider explicit kaon masses 
$M_{K^0}\neq M_{K^\pm}$. Assuming $\Lambda=$1 GeV we obtain the results 
listed in Table \ref{tab2}. Our result for the strong $f_0$ decay 
\eq 
\Gamma(f_0\to\pi\pi)=57.4\text{ MeV}\,,
\en
is consistent with the experimental data listed in Table \ref{tab:s1}.

\begin{table}[htbp]
\caption{Strong decay width $f_0(980)\to\pi\pi$: experimental data.} 
\label{tab:s1}
\begin{tabularx}{\linewidth}{XXXX}
\toprule\hline
Data &PDG~\cite{Amsler:2008zz}&BELLE~\cite{Mori:2006jj}
&\cite{Barberis:1999an}\\\hline
$\Gamma(f_0\rightarrow\pi\pi)$ [MeV]&$40-100$
&$51.3^{+20.8\,+13.2}_{-17.7\,-3.8}$&$80\pm10$\\\hline\bottomrule
\end{tabularx}
\end{table}
Further theoretical predictions are indicated in Table \ref{tab:s2}
\begin{table}[htbp]
\caption{Strong decay width $f_0(980)\to\pi\pi$: theoretical approaches.} 
\label{tab:s2}
\begin{tabularx}{\linewidth}{lXXXXXX}
\toprule\hline
Reference & ~\cite{Efimov:1993ei}&~\cite{Volkov:2000vy}
&~\cite{Anisovich:2002ij}&~\cite{Scadron:2003yg}
&~\cite{Celenza:2000uk}&~\cite{Oller:1998hw}\\\hline
Meson structure&$q\bar q$&$q\bar q$&$q\bar q$&$q\bar q$
&$q\bar q$&hadronic\\\hline
$\Gamma(f_0\rightarrow\pi\pi)$ [MeV]&20&28&52-58&53&56&18.2\\
\hline\bottomrule
\end{tabularx}
\end{table}
which, unfortunately, cover a large range of values, even for the 
same structure assumption. Again, the present situation for 
$\Gamma(f_0\to\pi\pi)$ does not allow for a clear statement 
concerning the $f_0$ structure.

For the strong $a_0\to \pi\eta$ decay we obtain
\eq 
\Gamma(a_0\to \pi \eta)=61.0\text{ MeV}
\en 
which also matches with the experimental results listed 
in Table \ref{tab:s3}. Here, the quarkonium models of \cite{Barnes:1985cy} 
and \cite{Scadron:2003yg} clearly deliver larger results compared to the 
molecular interpretation and data.
\begin{table}[htbp]
\caption{Strong decay width $a_0(980)\to\pi\eta$: 
data and theoretical approaches.} 
\label{tab:s3}
\begin{tabularx}{\linewidth}{l|XcXcXcX|XcXcX}
\toprule\hline
Reference &&\cite{Amsler:2008zz} &&\cite{Achard:2001uu}
&&\cite{Barberis:2000cx}&&&\cite{Barnes:1985cy}&&\cite{Scadron:2003yg}&\\ 
\hline
&&\multicolumn{5}{c}{experimental data}&&&$q\bar q$&&$q\bar q$&\\\hline
$\Gamma(f_0\rightarrow\pi\pi)$ [MeV]&&50-100&&$50\pm13\pm4$
&&$61\pm19$&&&225&&138&\\
\hline\bottomrule
\end{tabularx}
\end{table}
In the strong decay sector $f_0-a_0$ mixing also generates the 
isospin-violating decays $f_0\to \pi\eta$ and $a_0\to\pi\pi$. 
In the context of our approach we obtain the results
\eq
\Gamma(f_0\to \pi \eta)&=&0.57\text{ MeV}\,,\\
\Gamma(a_0\to \pi \pi)&=&1.59\text{ MeV}\,,
\en
which, since the processes are forbidden by isospin symmetry, are 
strongly suppressed compared to the dominant strong decays discussed above.

\section{Summary}\label{sec:summary} 

The present framework, where the scalars are assumed to be hadronic 
$K\bar K$ molecules, provides a straightforward and consistent 
determination of the decay properties, in particular the coupling constants 
and decay widths. The radiative decay properties of the $a_0$ and $f_0$ 
mesons have been studied comprehensively within a clear and consistent 
model for hadronic bound states. At the same time essential criteria such 
as covariance and full gauge invariance with respect to the electromagnetic 
interaction are satisfied. 

Despite that we deal with a rather simple model, it allows to study the 
influence of the spatial extension of the meson molecule and isospin 
violating mixing. The coupling of the hadronic bound state to the 
constituent kaons, including $f_0$-$a_0$ mixing effects, has been determined 
by the compositeness condition which reduces the number of free parameters 
to only one, the size parameter $\Lambda$. 

Our results for the electromagnetic decays ($a_0/f_0\to\gamma\gamma$ and 
$\phi\to\gamma a_0/f_0$) and, in addition, the strong decay widths 
($f_0\to\pi\pi$ and $a_0\to\pi\eta$) are analyzed with respect to $f_0$-$a_0$ 
mixing and finite size effects. 

We come to the conclusion that the hadronic molecule interpretation is 
sufficient to describe both the electromagnetic and strong $a_0$/$f_0$ 
decays, based on the current status of experimental data. Furthermore, 
the $f_0-a_0$ mixing strength could be determined by a precise measurement 
of the ratio of the charged and neutral $a_0$ meson decays. The $f_0-a_0$ 
mixing strength could deliver new insights into the contributions being 
responsible for isospin-violating mixing and the meson structure issue.

\begin{acknowledgments}

This work was supported by the DFG under Contracts No. FA67/31-1, FA67/31-2, 
and No. GRK683. This research is also part of the EU Integrated
Infrastructure Initiative Hadronphysics project under Contract
N0. RII3-CT-2004-506078 and the President Grant of Russia
``Scientific Schools''  No. 817.2008.2.

\end{acknowledgments}

\appendix

\section{Loop Integrals}

Here we give a short presentation of the structure integrals and its 
evaluation relevant for the derivation of the transition form factors. 
For simplicity we restrict to the diagrams of Figs. \ref{fig4}~(a,b) 
and \ref{fig5}~(a,b), which do not contain contact 
vertices. The additional diagrams generated due to nonlocal effects are 
discussed in detail in \cite{Branz:2007xp,Faessler:2003yf}. The full 
structure integrals characterizing the electromagnetic decays are given by 
\bea
I_{S\gamma V}^{\mu\nu}(M_S^2,0,M_V^2)&=&\int\frac{d^4k}{\pi^2i}\,
\widetilde\Phi(-k^2)\left((2k+p-q)^\mu(2k-q)^\nu S_K\big(k+\dfrac p2\big)
S_K\big(k-\dfrac p2\big)S_K\big(k+\dfrac p2-q\big)\right.\nonumber\\
&&\left.+g^{\mu\nu}S_K\big(k+\dfrac p2\big)S_K\big(k-\dfrac p2\big)\right)
\,,\label{2}\\
I_{\phi S\gamma}^{\mu\nu}(M_\phi^2,M_S^2,0)&=&\int\frac{d^4k}{\pi^2i}\,
\widetilde\Phi(-k^2)
\left(\frac{(2k-q-p)^\nu(2k-q)^\mu}{S_K\big(k+\dfrac p2\big)
S_K\big(k-\dfrac p2\big)S_K\big(k-\dfrac p2-q\big)}
+\frac{g^{\mu\nu}}{S_K\big(k+\dfrac p2\big)S_K\big(k-\dfrac p2\big)}
\right)\,,\nonumber
\ena
where $q$ is the photon momentum and $p$ of the scalar. 
In the case of the two-photon decay the expressions corresponding to all 
the diagrams of Fig. \ref{fig4} are quoted in \cite{Branz:2007xp}. 
We use the expression for the $S\to V\gamma$ decay (Eq.~\ref{2}) as an 
example to demonstrate the technique for the derivation of the loop 
integral $I_{S\gamma V}(M_S^2,0,M_V^2)$. In the first step we separate 
the gauge invariant part of the full expression $I^{\mu\nu}$ by writing
\eq
I_{S\gamma V}^{\mu\nu}(M_S^2,0,M_V^2)=I_{S\gamma V}(M_S^2,0,M_V^2)
b^{\mu\nu}+I^{(2)}_{S\gamma V}(M_S^2,0,M_V^2)c^{\mu\nu}
+\delta I_{S\gamma V}\,,\label{eq:i}
\en
where the remainder term $\delta I_{S\gamma V}$ contains the noninvariant 
terms. The tensor structures $b^{\mu\nu}$ and $c^{\mu\nu}$ have already 
been defined in (\ref{eq:b}). Since we deal with real photons, only the 
first term of (\ref{eq:i}), proportional to $b^{\mu\nu}$, is relevant. 
In the second step Feynman parametrization is introduced and the 
integration over the four-momentum $k$ is performed. For instance, 
in the local limit we obtain
\eq 
I_{S\gamma V}(M_S^2,0,M_V)&=&\int\limits_0^1d^3\alpha\,
\delta(1-\sum\limits_i\alpha_i)\,
\frac{4\alpha_1\alpha_3}{M_K^2-M_S^2\alpha_1\alpha_3-M_V^2\alpha_2\alpha_3}\,.
\en 
The mathematical treatment of the diagrams including contact vertices 
is straightforward and in complete analogy with the above example.

The loop integrals of the diagrams contributing to the strong decays 
(Fig. \ref{fig6} (a) and (b)) read as
\eq 
I^{(a)}(M_K^2,p^2,q_1^2,q_2^2)&=&\frac{g_\pi g_{\pi(\eta)}}{(4\pi)^2}
\int\frac{d^4k}{\pi^2i}\,\widetilde\Phi(-k^2)
\big(k-\frac p2-q_2\big)_\mu(k+\frac p2+q_1)_\nu S_K\big(k+\frac p2\big)
S_K\big(k-\frac p2\big)S^{\mu\nu}_{K^\ast}\big(k+\frac p2-q_1\big)\,,
\nonumber\\
I^{(b)}(M_K^2,p^2,q_1^2,q_2^2)&=&-\frac{1}{M_{K^\ast}^2}
\frac{g_\pi g_{\pi(\eta)}}{(4\pi)^2}\int\frac{d^4k}{\pi^2i}\,
\widetilde\Phi(-k^2)\big(k-\frac p2-q_2\big)(k+\frac p2+q_1) 
S_K\big(k+\frac p2\big)S_K\big(k-\frac p2\big)\,.
\en 
Again, we evaluate the above expressions by introducing Feynman parameters 
and integrating over the loop-momentum~$k$. 

\section{Gauge invariance}

In this appendix gauge invariance is demonstrated by means of the charged 
$a_0$ meson decays. The kaon-loop integral corresponding to the 
diagrams (a) and (b) of Fig. \ref{fig5} is given by
\eq 
I^{\mu\nu}_\bigtriangleup&=&\int\frac{d^4k}{\pi^2i}\widetilde\Phi(-k^2)
\Big\{S\Big(k+\frac p2\Big)S\Big(k-\frac q2\Big)
S\Big(k-\frac p2\Big)(2k+q_2)^\mu(2k-q_1)^\nu\nonumber\\
&+&g^{\mu\nu}S\Big(k+\frac p2\Big)S\Big(k-\frac p2\Big)\Big\}\,,
\en 
where $q=q_1-q_2$. The part $I^{\mu\nu}_{\bigtriangleup_\perp}$ being 
gauge invariant with respect to the photon momentum $q_1^\mu$ is separated 
from the so-called remainder term $\delta I^{\mu\nu}_\bigtriangleup$ by using
\eq 
(2k+q_2)^\mu&=&(2k+q_2)^\mu_{\perp q_1}+q_1(2k+q_2)\frac{q_1^\mu}{q_1^2}
\nonumber\\
g^{\mu\nu}&=&g^{\mu\nu}_{\perp q_1}+\frac{q_1^\mu q_1^\nu}{q_1^2}\,.
\en 
Therefore, the noninvariant term is given by
\eq
\delta I^{\mu\nu}_\bigtriangleup&=&\int\frac{d^4k}{\pi^2i}
\widetilde\Phi(-k^2)\Big\{\Big[S\Big(k+\frac p2\Big)
S\Big(k-\frac p2\Big)-S\Big(k-\frac p2\Big)S\Big(k-\frac q2\Big)\Big]
\frac{q_1^\mu}{q_1^2}(2k-q_1)^\nu\nonumber\\
&+&S\Big(k+\frac p2\Big)S\Big(k-\frac p2\Big)
\frac{q_1^\mu q_1^\nu}{q_1^2}\Big\}\nonumber\\
&=&-\int\frac{d^4k}{\pi^2i}\widetilde\Phi(-k^2)S\Big(k-\frac p2\Big)
S\Big(k-\frac q2\Big)\Big]\frac{q_1^\mu}{q_1^2}(2k-q_1)^\nu\,.
\en 

For the bubble diagram (c) of Fig. \ref{fig5} the loop integral reads 
as (see \cite{Faessler:2003yf})
\eq 
I^{\mu\nu}_{bub}=-\int\frac{d^4k}{\pi^2i}
\big(2k+\frac{q_1}{2}\big)^\mu k^\nu\int\limits_0^1dt
\widetilde\Phi^\prime\big[-\big(k+\frac{q_1}{2}\big)^2t-k^2(1-t)\big]\,.
\en 
This leads to the remainder
\eq 
\delta I^{\mu\nu}_{bub}=\int\frac{d^4k}{\pi^2i}\widetilde\Phi(-k^2)
\frac{q_1^\mu}{q_1^2}(2k-q_1)^\nu S\Big(k-\frac p2\Big)S\Big(k-\frac q2\Big)
\en 
which cancels with $\delta I^{\mu\nu}_\bigtriangleup$ and therefore
\eq 
\delta I^{\mu\nu}=\delta I^{\mu\nu}_{\bigtriangleup}
+\delta I^{\mu\nu}_{bub}=0\,.
\en

\section{Summary Table}
For completeness we indicate in the following tables the full list 
of couplings and transition widths for electromagnetic and strong decays.
\begin{table}[thpb]
\caption{$f_0$ and $a_0$ decay properties with and without $f_0-a_0$ 
mixing for local and nonlocal ($\Lambda=1$ GeV) interaction.}
\label{tab1}
\begin{tabularx}{8.8cm}{|l|l|XX|XX|}
\hline
&&\multicolumn{2}{c|}{Without mixing}&\multicolumn{2}{c|}{With mixing}\\
&&local&nonlocal&local&nonlocal\\\hline
\multirow{2}{*}{$f_0\to\gamma\gamma$}&$g$ [GeV$^{-1}$]
&0.086&0.079&0.085&0.078\\
&$\Gamma$ [keV]&0.29&0.25&0.29&0.24\\\hline
\multirow{2}{*}{$f_0\to\rho\gamma$}&$g$ [GeV$^{-1}$]&0.425&0.411&0.421&0.407\\
&$\Gamma$ [keV]&8.09&7.58&7.93&7.44\\\hline
\multirow{2}{*}{$f_0\to\omega\gamma$}&$g$ [GeV$^{-1}$]
&0.431&0.418&0.427&0.414\\
&$\Gamma$ [keV]&7.57&7.12&7.43&6.99\\\hline
\multirow{2}{*}{$\phi\to f_0\gamma$}&$g$ [GeV$^{-1}$] &1.97&&1.95&\\
&$\Gamma$ [keV]&0.64&&0.63&\\\hline
\end{tabularx}
\hfill
\begin{tabularx}{8.85cm}{|l|l|ll|ll|}
\hline
&&\multicolumn{2}{c|}{without mixing}&\multicolumn{2}{c|}{with mixing}\\
&&local&nonlocal&local&nonlocal\\\hline
\multirow{2}{*}{$a_0\to\gamma\gamma$}&$g$ [GeV$^{-1}$] &$0.076\quad\quad\,$
&0.069&$0.080\quad\quad\;$&0.073\\
&$\Gamma$ [keV]&0.23&0.19&0.26&0.21\\\hline
\multirow{2}{*}{$a_0\to\rho\gamma$}&$g$ [GeV$^{-1}$]&0.388&0.372&0.408&0.391\\
&$\Gamma$ [keV]&7.18&6.59 &7.94&7.29\\\hline
\multirow{2}{*}{$a_0\to\omega\gamma$}&$g$ [GeV$^{-1}$]
&0.394&0.378&0.414&0.398\\
&$\Gamma$ [keV]&6.76&6.22 &7.47&6.88\\\hline
\multirow{2}{*}{$\phi\to a_0\gamma$}&$g$ [GeV$^{-1}$]&1.82&&1.91&\\
&$\Gamma$ [keV]&0.37&&0.41&\\\hline
\end{tabularx}
\end{table}
\begin{table}[thbp]
\caption{Strong $a_0$ and $f_0$ decay properties.}
\label{tab2}
\begin{tabularx}{8.5cm}{|l|X|X|}
\hline
&$g$ [GeV]&$\Gamma$ [MeV]\\\hline
$f_0\to\pi\pi\quad$&1.40&57.4\\\hline
$a_0\to\pi\eta\quad$&2.15&61.0\\\hline
\end{tabularx}
\hfill
\begin{tabularx}{8.5cm}{|l|X|X|}
\hline
&$g$ [GeV]&$\Gamma$ [MeV]\\\hline
$f_0\to\pi\eta\quad$&0.208&0.57\\\hline
$a_0\to\pi\pi\quad$&0.234&1.59\\\hline
\end{tabularx}
\end{table}

\end{document}